\documentclass[12pt,a4paper]{article}
\usepackage{color}
\usepackage{amssymb}
\usepackage{times}
\usepackage{graphicx}
\usepackage{setspace}
\usepackage[latin2]{inputenc}

\newif\ifAMStwofonts

\newcommand{\bc}{\begin{center}}
\newcommand{\be}{\begin{equation}}
\newcommand{\ee}{\end{equation}}
\newcommand{\ec}{\end{center}}

\newcommand{\spose}[1]{\hbox to 0pt{#1\hss}}
\newcommand{\lta}{\mathrel{\spose{\lower 3pt\hbox{$\mathchar"218$}}
 \raise 2.0pt\hbox{$\mathchar"13C$}}}
\newcommand{\gta}{\mathrel{\spose{\lower 3pt\hbox{$\mathchar"218$}}
 \raise 2.0pt\hbox{$\mathchar"13E$}}}

\definecolor{grayblue}{cmyk}{0.20,0.15,0.0,0.1}
\definecolor{darkblue}{cmyk}{0.60,0.50,0.0,0.50}
\definecolor{vdarkblue}{cmyk}{0.80,0.75,0.0,0.55}
\definecolor{lightsea}{cmyk}{0.15,0.05,0.0,0.15}

\begin{document}

\begin{center}
{\bf \Large Nature of light variations in the symbiotic binary V417 Cen}
\end{center}
\begin{center}
M. Gromadzki$^{1}$, J. Miko{\l}ajewska$^{1}$, 
B. Pilecki$^{2}$, P. Whitelock$^{3,4}$, M. Feast$^{3,4}$
\end{center}
\begin{center}
$^{1}$Nicolaus Copernicus Astronomical Center, Bartycka 18, 00-716 Warsaw, Poland\\
$^{2}$Warsaw University Observatory, Al. Ujazdowskie 4, 00-478 Warsaw, Poland\\
$^{3}$South African Astronomical Observatory, PO Box 9, 7935 Observatory, South Africa\\
$^{4}$Astronomy Department, University of Cape Town, 7701 Rondebosch, South Africa\\
\end{center}

\vspace{0.2cm}

\large
\vspace{.6cm} 
\hspace{-2cm}
\begin{center}
{\bf  Abstract }
\end{center}

V417 Cen is a D'-type symbiotic system surrounded by a faint, extended
asymmetric nebula. Optical photometric observations of this object cover 
last 20 years. They show strong long term modulation with a period of about 
1700 days and amplitude about 1.5 mag in V band, in addition to variations 
with shorter times--cales and much lower amplitudes. In this presentation we discuss possible reasons of these variations.

\vspace{0.5cm}



\section{Introduction}

Despite the fact that V417~Cen is unique even for a D'-type symbiotic star
(only seven such objects are known) and relatively bright ($V \sim 12$ mag) it is poorly explored.
The symbiotic nature of this object was proposed by Steiner et al. (1988) but most facts about this system one can find in Van Winckel et al. (1994).
For the cool companion they derived G2Ib--II,
$\log{L/L_{\odot}}=3.5$, $T_{eff}=5000$ K and $\log{g} = 1.5\pm 0.5$.
This implies $R_{\rm g}= 75 R_{\odot}$ and $M_{\rm g}= 5-7 M_{\odot}$. 
V417~Cen is associated with  a ring nebula ($r \approx 0.4 {\rm pc}$,
assuming $d \approx 5 {\rm kpc}$), and more extended ($\approx 2.5 {\rm pc}$) 
remnants of bipolar nebula (Van Winckel et al. 1994). Zamanov et al. (2006) estimated $v_{\rm rot} \sin(i) = 75 {\rm km~s}^{-1}$,
what is 71 \% of critical value and implies very short rotation period
$P_{\rm rot} \lesssim 50.6\pm10.2$ d.

\section{Photometry}

We have collected $V$-band ASAS photometry (Pojmanski et al. 2002) 
and visual AAVSO observations. The data cover 2000-2010.
A few measurements from 1986--1993 are found in Cieslinski et al (1994, 1997). 
The $V$-band light curve is shown in Fig. 1, together with $V-I$, $J-K$ colours. 
$V-I$ colours are taken from ASAS. $J-K$ colours are taken from different sources: Van Winckel et al.
(1994), 2MASS, DENIS and our observations from SAAO. The colours are transformed
to SAAO photometric system using Carpenter (2001) equations. 

\section{Period analysis}

A period search, based on the modified Lomb--Scargle method (Press \& Rybicki, 1989), was carried out. Because amplitude of variation is significant $\Delta V = 2.3$ mag, magnitudes are converted to flux. The period analysis of ASAS+AAVSO set of data resulted in a period of $P_{1}=1698\pm220$ days. Sine curve fitted to this data extended by measurements from Cieslinski et al. (1994,1997) gave slightly shorter period of $1652\pm11$ days.
The power spectrum of residual light curve shows $P_{2}=582\pm24$  
days. After removing from the light curve this period,  
the power spectrum become more complicated and shows series of
peaks corresponding to several periods. 
The strongest one is $P_{3}=278\pm6$ d. It is worth to notice
that $P_{1} \approx 3 P_{2} \approx 6 P_{3}$. Light curves folded with the three strongest periods and the corresponding power spectra are plotted on Fig.~2. 
These results do not agree with historical photographic observations.
Van Winckel et al. (1994) found a 246-day periodicity with an amplitude of 0.5 mag, using magnitudes estimated on 238 Harvard plates, taken between 1919 and 1934, and on 3 Sonneber plates taken 1959.
They interpreted this periodicity as orbital modulation due to reflection effect.

\section{Spectrum}

Emission spectrum of V417~Cen is very untypical and variable. Only few emission lines are present: [O {\sc iii}]
$\lambda$4959, $\lambda$5007, H$\alpha$, H$\beta$, He {\sc i} $\lambda$5876, [N {\sc ii}] $\lambda$6548, $\lambda$6584. [O {\sc iii}] $\lambda$5007 
is usually the strongest emission line whereas H$\alpha$ is relatively faint and sometimes absent (Steiner et al., 1988; Cieslinski et al., 1994, Van Winckel et al., 1993, 1994; Munari \& Zwitter, 2002).

The spectrum variation are best illustrated in Van Winckel et al. (1994). They presented two low
resolution spectra obtained in Feb 1988 and Jan 1993. In 1988 the continuum appeared to be stronger  than in 1993 by
a factor 3. In 1993 H$\alpha$ was relatively strong, and its flux was equal to about 2/3 
of flux in [O {\sc iii}] $\lambda$5007. In Feb 1988  H$\alpha$ was very faint, and in Jul 1988 was absent. 
Such a behaviour may suggest that the continous spectrum is dominated by the cool component: the emission lines appear to become relatively stronger as the continuum decreases.

Another low resolution spectra obtained in Apr 1995 and May 1996 were published
by Munari \& Zwitter (2002). In 1995 the continuum appeared to be a factor of 2 fainter than in spectrum obtained in Feb 1988  by Van Winckel et al. (1994) but emission lines looked similar.
In 1996 the continuum became slightly fainter ($\sim$70 \%) but intensity of [O {\sc iii}] dramatically decreased and seemed similar to H$\alpha$.   
This suggests that the intrinsic emission line fluxes are poorly correlated with continuum changes. We suspect that emission lines are produced mainly by the asymmetric nebula and their intensity may depend on slit orientation.

Zamanov et al. (private communication) obtained 8 high resolution spectra during about 100-day period in 2004 (close to the minimum of the 1700-day period).
They found strong variability of emission lines with time--scales of about 1 month, and no variability was detected with time--scales of day.

\section{Discussion and conclusions}

Based on the available observations of V417 Cen, it is hard to indicate
unambiguously causes of its variability. The light curve is too short with respect to the longest period found ($P_{1}=1700$ d), and it is even impossible to confirm its coherency. The measurements from Cieslinski et al. (1994,1997) suggest that it is not coherent. The continuum changes observed in the spectra obtained by Munari \& Zwitter (2002) do not agree with those predicted by our fitted sine curve.

We cannot exclude short orbital periods ($P_{2} = 580$ d and $P_{3} = 280$ d) because time-scale of synchronization is relatively long and comparable with lifetime of symbiotic stars ($\tau_{\rm syn} \sim 10^5$ yr). However objects surrounded by extended bipolar nebulae usually have long orbital period, of order of dozens of years, and they are symbiotic Miras. Zamanov et al. (2006) suggested that the system is synchronized and the orbital period is equal to $\sim$50 d. We can rather reject such a short period  because in this case the cool giant would fill its Roche lobe and the system should be very active. 

We expect a long orbital period. The period $P_{1} = 1700$d is the longest detected and we assume that it is the orbital period
of V417~Cen. In this case, the light changes could be produced by several mechanisms.
We can exclude a modulation due to reflection effect, because of 
the high amplitude of variations. Even systems with more active hot components and smaller separations do not show such large amplitudes.

In symbiotic binaries two kinds of light variations caused by accretion rate changes can be observed.     
The first one is due to enhanced accretion during periastron passage in systems with elliptical orbit. 
System with long period tend to have eccentric orbits (eg. BX~Mon, MWC~560, CD-43 14 304) but a weak point of this explanation is strong tidal interaction, which relatively fastly slows down rotation of the cool component.
The second possibility is accretion instability, which is, for example, observed in the symbiotic recurrent novae RS~Oph and T CrB between their TNR-nova outbursts.
The time-scales of such activity (1000--2000 d) well fit the time-scales observed in V417~Cen. 
Unfortunately, in both cases the hot component should dominate the $V$-band light, just opposite to what is suggested by the spectroscopic observations of V417 Cen.

We propose therefore an alternative model -- obscuration and reflection of the cool giant radiation caused by gas/dust accumulated
in the binary system mainly around the hot companion and in $L_{4}$ and $L_{5}$ Lagrange points. Such solution have a few advantages:

\begin{itemize}
\item Long synchronization time-scales ($\tau_{\rm syn} = 5 \times 10^{7}$ yr assuming $P_{orb} \approx 1700$ d), so tidal interactions do not slow down the rotation of the donor star.

\item Obscuration by gas/dust around the hot component and $L_{4}$ and $L_{5}$ Lagrange points explains presence of $P_{2}=P_{1}/3$ as three minima should be observed during orbital cycle. Light curve shows two shallow minima around mid 2002 and 2005 (maybe caused by mater in Lagrange points) and a deeper one in 2004 (maybe caused by the material around the hot component).    

\item  Assuming $P_{orb} \approx 1700$ d, $M_{\rm g}= 5-7 M_{\odot}$ and $M_{\rm g}= 0.5-1 M_{\odot}$ results in a separation between the components equal to $\sim 5$ AU. It is hard to explain presence of
dust so closely to binary components, although Angeloni et al. (2007) explained the observed spectral energy distribution of another D'-type symbiotic star -- HD 330036 -- 
by presence of a few dust shells, with the closest one having a radius of $\sim 2$ AU.   

\item Reflection off dust could explain continuum variation of the cool star, which
 should be the brightest when it is in front of its hot companion and most of dust is located at
opposite side with respect to observer.

\item  $V-I$, $J-K$ colour variations correspond to the reflection off dust. The object is bluer in maximum and redder in minimum.
\end{itemize}

Unfortunately, the long period seems to be noncoherent and the nature of light variations and the length of the orbital period
remain still unknown. Further observations are necessary to fully understand nature of V417~Cen. 
\\\\

{\it Acknowledgments}. This study made use of American Association of Variable Star Observers (AAVSO) International Database contributed by worldwide which we acknowledged. The project was partly supported by the Polish Research Grant No. N203 395534.


\begin{figure}
\begin{spacing}{1.8} 
\includegraphics[width=16cm]{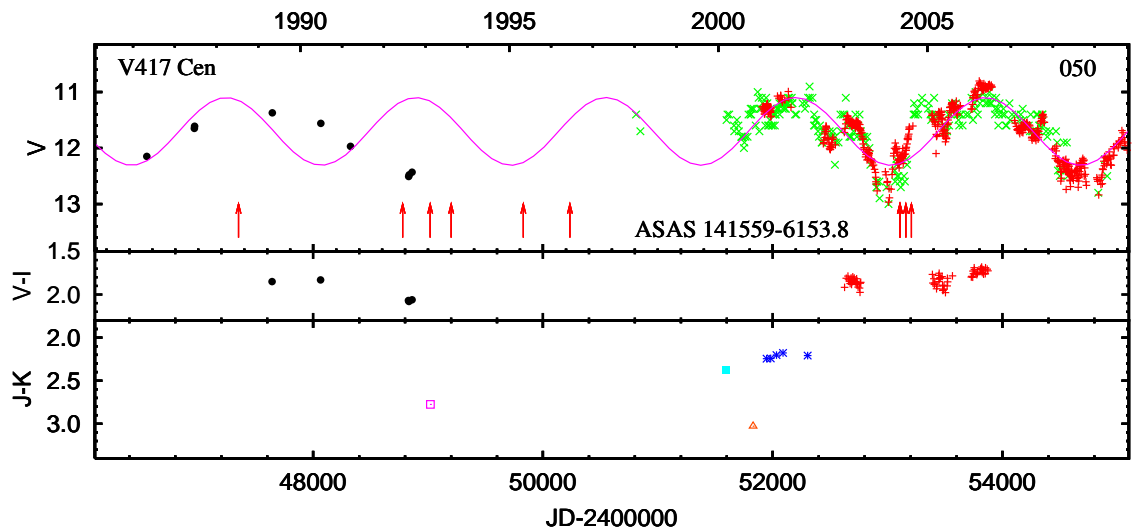}
\caption{\normalsize \label{mig100} 
The $V$-band/visual light curve of V417 Cen. Red crosses represents ASAS data,
green times AAVSO data and black dots are taken from Cieslinski et al. (1994,1997). Blue asterisks represents SAAO data, cyan filled rectangle 2MASS data,
red empty triangle DENIS and magenta empty rectangle is taken from Van
Winckel et al. (1994). Red arrows indicate moments of spectroscopic observations (see Sec. 4).
}
\end{spacing}
\end{figure}

\begin{figure}
\begin{spacing}{1.8} 
\includegraphics[width=10cm]{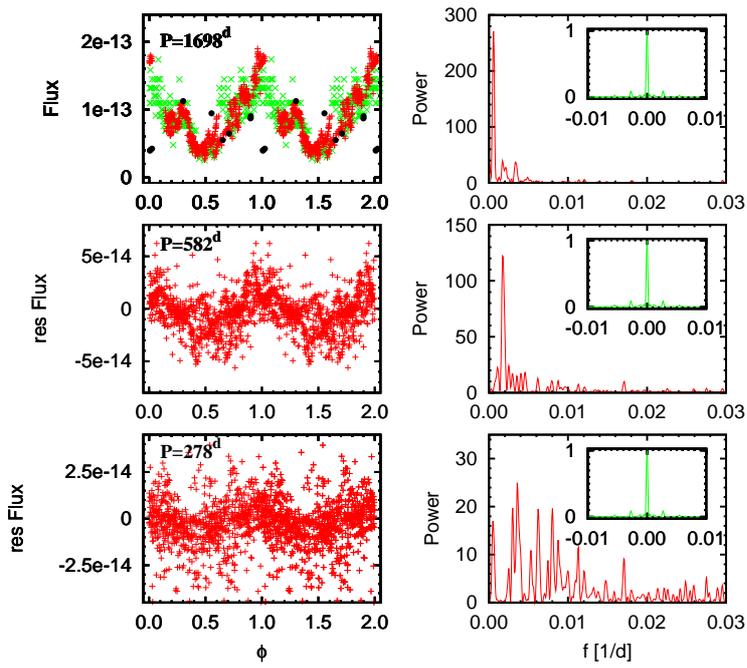}
\caption{\normalsize \label{mig100}
Light curves of V 417 Cen folded with the strongest periods found (left) and corresponding power spectra (right).}
\end{spacing}
\end{figure}

\end{document}